\documentstyle[floats,prl,aps]{revtex}
\begin{document}
%\draft
\title{Exact first-passage
exponents of 1D domain growth: relation to a reaction-diffusion model}

\author{Bernard Derrida$^{\dag \ddag}$, Vincent Hakim$^{\dag}$ and
Vincent Pasquier$^{\ddag}$}
\address{$\dag$ Laboratoire de Physique Statistique,
 24 rue Lhomond, 75231 Paris Cedex 05, France.\\
$\ddag$Service de Physique Th\'{e}orique
CE Saclay,F91191 Gif sur Yvette, France.}

\author{\parbox{397pt}{\vglue 0.3cm \small
In the zero temperature Glauber dynamics of the ferromagnetic Ising or
$q$-state Potts model,
the size of domains is known to grow like $t^{1/2}$. Recent simulations have
shown that the fraction $r(q,t)$ of spins which have never
flipped up to time $t$ decays like a power law $r(q,t) \sim t^{-\theta(q)}$
with a non-trivial dependence of the exponent  $\theta(q)$
on $q$ and on space dimension.
By mapping the problem on an exactly soluble
 one-species coagulation model ($A+A\rightarrow
A$), we obtain the exact expression of $\theta(q)$ in dimension one.
}}

\maketitle

\pacs{02.50.+s, 05.40.+j, 82.20-w}

VERSION :\today

Phase ordering and domain growth in systems quenched from a
disordered phase to an ordered phase has been a subject of much interest
during the last fifteen years in fields ranging from metallurgy
to cosmology \cite{revart}. It is well established that the
pattern of growing domains is self-similar in time and that the characteristic
domain size increases with a simple power law $t^{\rho}$. For example,
$\rho=1/2$ holds for all systems with short range-interactions
described by a scalar non-conserved order parameter. However,
as noted for the
 auto-correlation function \cite{furu},
 correlations at different times  are characterized by more complicated
exponents.
 Recently,
the fraction of spins $r(q,t)$
which have never flipped up to time $t$ has been
measured in  simulations of coarsening at zero-temperature for the
Ising and  for the
$q$-state Potts models. This fraction decreases with time like a power law
\cite{DBG,Stauf},
\begin{equation}
r(q,t) \sim t^{-\theta(q)}
\label{frac1}
\end{equation}
and   numerical data indicate that the exponent $\theta(q)$ varies  both with
$q$ and the dimension of space. Since $dr(q,t)/dt$ measures
the probability that a given point is crossed for the first time at time  $t$
by a
domain wall,
$\theta(q)$ can
 be viewed as a  first passage
exponent \cite{Cardy}.

The aim of the present letter is to give the exact expression
of $\theta(q) $ in one dimension.
\begin{equation}
\theta(q)= -\frac{1}{8} +\frac{2}{\pi^2} \left[\cos^{-1}
 \left(\frac{2-q}{\sqrt{2}\  q}\right) \ \right]^2
\label{mform}
\end{equation}
This result fully agrees with previous numerical predictions
based on MonteCarlo simulations \cite{DBG,Stauf} or on finite size scaling
calculations
\cite{der1}.
It implies that for  the Ising model $\theta(2)=3/8$ is exact.  Note however
that for other
choices of $q$,  the exponent $\theta(q)$ is in general irrational (for example
$\theta(3)=
.53795082..$).

To obtain (\ref{mform}), we  are going to follow four main steps:
first by   using finite size scaling we will  relate  the exponent $\theta(q)$
to the large $L$ behavior of the  fraction
$\rho_L(q)$ of spins which never flip between
 time $0$ and time $\infty$ for a finite
  one dimensional system of $L$ sites with periodic
boundary conditions (for the zero temperature
Glauber dynamics of the $q$-state Potts model);
secondly, we will show that    the calculation of $\rho_L(q)$
can be reduced to solving the steady state of
 a reaction-diffusion model
($A+A \to A$) on a one dimensional lattice of $L$ sites with
a source  of particles at the origin (i.e. at site $0 \equiv L$);
our third step will be  the  solution of the steady state of that
reaction-diffusion model
%(on a finite lattice of $L$ sites  with the source at the origin)
leading to the exact expression of $\rho_L(q)$ for arbitrary $L$ and $q$;
lastly, we will extract the exponent $\theta(q)$ from the  large $L$
behavior of this expression.

In an infinite system, it is known \cite{Racz,Bray1d} that
starting with a random initial condition, the size of domains grows like
$t^{1/2}$.
For a finite system of size $L$, one expects the dynamics to be very much the
same as for the infinite system when $t \ll L^2$
(as the size of domains is small compared to the system size).
On the other hand, for $t \gg L^2$, there
is a single domain left in the system and the dynamics stops. Therefore one
expects that $\rho_L(q) \sim r(q,L^2)$ which implies that
\begin{equation}
\rho_L(q) \sim L^{-2 \theta(q)}
\label{fss}
\end{equation}
We  obtain, in what follows, an exact expression of
$\rho_L(q)$   valid for all sizes $L$ and (\ref{fss}) will allow us to extract
then the exponent $\theta(q)$.

We  now  show that the calculation of $\rho_L(q)$
is equivalent to solving  the following reaction-diffusion \cite{reactdif}
model  ($A+A\rightarrow A$) defined on a ring of $L$ sites with a source of
particles at the origin (that we  choose to be at site $L\equiv 0$):
 the origin (site $L$) is always occupied and
the other  sites   $i$  (for $ 1 \leq i \leq L-1$) are either
occupied by a particle $A$ or empty; during every
 infinitesimal time interval $dt$, each particle hops with probability $dt$ to
its right neighbor  and  with probability $dt$ to its left neighbor (and does
not move with probability $1-2dt$); if two particles occupy the same site, they
instantaneously coagulate ($A+A \to A$); in addition
 whenever the particle  at the
origin
jumps to one of its neighbors, a new particle is  instantaneously produced at
the origin.

In the steady state, injection of particles at the origin is balanced by
aggregation
of particles in the bulk and one has a probability
$P_L(m)$ of finding $m$ sites occupied on the ring of $L$ sites. The connection
with the
spin problem described above is made \cite{der1} by expressing
the probability $\rho_L (q)$ of never flipping
(from $t=0$ until $t= \infty$) in terms of $P_L (m)$
\begin{equation}
 \rho_L (q) = \sum_{m=1}^L \ P_L (m) \ {1 \over q^{m-1}}  \ \ .
\label{relat}
\end{equation}

This formula is obtained by remembering that updating a spin in one
dimension  with Glauber dynamics  at zero temperature simply consists in
choosing
for its new value the value of one of its two neighbors at random. When the
value $V$ of a spin $S_i$ at time t is traced back in time, a random walk
is obtained which connects $S_i (t)$ through various ancestors to a
particular spin in the initial configuration with value $V$. Now, let us
consider all the updates of the spin at the origin. They all give rise to
random walks going backward in time. These walks
can merge and are created at the origin exactly as in the
one species coagulation model. So, in the limit $t \to \infty$,  they lead to
$m$
different ancestors in the initial configuration with probability
$P_L (m)$. Equation (\ref{relat}) then follows by noting that the
spins in the  initial configuration are
uncorrelated and that the spin at the origin  never flips if
and only if all its
updates have the same value.

%In the following, we firstly generalize
%existing results on the one species coagulation
%model so as to obtain an exact explicit
%expression for $\rho_L (q)$ (eq.(\ref{sum},\ref{det1}) below). We then study
%%its
%asymptotic behavior for large $L$ and derive the formula (\ref{mform}) for
%$\theta(q)$.

We now come to the full exact solution of the reaction-diffusion model
steady state which is  the crucial point
of the present work.
It is  known that some properties of the coagulation model
$A+A \to A$ can be calculated exactly \cite{Bram} even in the presence of a
fixed source \cite{source}.
The properties which have been calculated  \cite{Bram}  so far on this sort of
problems are the probabilities that a (connected) region   of consecutive sites
are all empty. This is because one
can obtain closed kinetic  equations   (see (\ref{dif}) below) for these
quantities.
The key to our exact solution is that more complicated quantities like the
probability of having 2 (or 3 or $\cdots n$)   disconnected  empty regions
 can be expressed in a simple way in terms of these
probabilities that a single  connected region is empty.

Let us define for the coagulation model
the probabilities $B_{i,j},\ 1\leq i<j\leq L$ that
the segment $\{i,i+1,...,j-1\}$ contains no
particle \cite{Bram,der1}. These quantities change when a particle
enters or leaves the segment through its extremities and this can be expressed
in terms of the $B_{i,j}$ themselves (for example, the probability to find a
particle at
site $j$  given that sites $i,i+1, \cdots,j-1$ are empty is
$B_{i,j}-B_{i,j+1}$).
Therefore, the $B_{i,j}$'s
satisfy closed kinetic equations which read in the steady state
\begin{equation}
B_{i+1,j}+B_{i-1,j}+B_{i,j+1}+B_{i,j-1} - 4 B_{i,j} = 0
\label{dif}
\end{equation}
Modified equations for $j=i+1$ and
boundary conditions coming from the permanent occupation of the origin
can be taken into account by the conventions $B_{i,i}=1$ and
$B_{0,j}=B_{i,L+1}=0$.

 The explicit solution of   (\ref{dif}) is
\cite{der1}:
\begin{equation}
B_{i,j}= \sum_{k \ even}  \ \sum_{k' odd}\frac{
8 \sin k \alpha  \  \sin k' \alpha \ ( \sin k i \alpha \ \sin k' j \alpha  \
- \ \sin kj \alpha  \ \sin k' i \alpha )}{
(L+1)^2 \ (\cos k' \alpha \ - \ \cos k \alpha ) \ ( 2 - \cos k \alpha - \cos k'
\alpha ) }\ ,\ \ i<j
\label{Bsol}
\end{equation}
 with $
 \alpha =\pi/(L+1),\ 2 \leq k \leq L,\ 1 \leq k' \leq L$
 (Note that (\ref{Bsol}) is not valid for $j=i$ since it gives $B_{ii}=0$
instead of $1$).
 For example, when $L=4$, this gives
$B_{1,2}= B_{3,4}=26/44; B_{1,3}= B_{2,4}=16/44; B_{1,4}=8/44;
B_{2,3}= 30/44$.

These known results can be
generalized by introducing the probabilities $B^{(n)}_{i_1,i_2,\cdots
,i_{2n-1},i_{2n}}$
that there is no particle in any of the disconnected segments
$\{i_1,i_1+1,\cdots,i_2-1\},\cdots \{i_{2n-1},\cdots,i_{2n}-1\}$ with
$i_1<i_2<\cdots<i_{2n}$. Similarly to the
$B_{i,j}$'s (to which they reduce for $n=1$), the $B^{(n)}$'s satisfy
equations analogous to (\ref{dif}). The new boundary conditions can be taken
care of
by the convention that
$B^{(n)}$
 reduces to the corresponding $B^{(n-1)}$ if
two successive indices coincide.

It turns out that the diffusion-like equations satisfied by the $B^{(n)}$'s
 can be solved exactly and that their solutions
 can all be expressed in terms of the $B_{i,j}$. Namely,
\begin{equation}
B^{(2)}_{i,j,k,l} = B_{i,j}B_{k,l}+B_{i,l}B_{j,k}-B_{i,k}B_{j,l}\
\label{B2ij}
\end{equation}
and more generally
\begin{equation}
B^{(n)}_{i_1,i_2,...,i_{2n-1},i_{2n}}=\frac{1}{2^n \ n!}\sum_{\sigma}
\epsilon(\sigma) B_{i_{\sigma(1)},i_{\sigma(2)}} \cdots
B_{i_{\sigma(2n-1)},i_{\sigma(2n)}}
\label{Bnij}
\end{equation}
where the sum runs over all the permutations of the
indices $\{i_1,i_2,\cdots,i_{2n}\}$ and
$\epsilon(\sigma)$ is the signature
of the permutation $\sigma$ (Note that (\ref{B2ij}) and (\ref{Bnij})
 are Pfaffians \cite{meh}).
One  way to prove  (\ref{B2ij}-\ref{Bnij})  is simply to check that
their r.h.s. satisfy the same diffusion-like equations as the $B^{(n)}$'s
when $B_{i,j}$ is solution of (\ref{dif}) and that they indeed reduce
to the expression for $B^{(n-1)}$ if two successive indices coincide.
In the case $L=4$, one finds that $B^{(2)}_{1,2,3,4}= 15/44$.

Once  the $B^{(n)}$'s are known, all the steady state properties
of the reaction-diffusion model are in principle computable, in particular  the
$P_L(m)$,   and as a consequence (\ref{relat}) the $\rho_L(q)$.
There is a simple way of obtaining the $\rho_L(q) $  in terms of the
$B^{(n)}$'s:
introduce the (normalized) weights $w(\tau_{1},\cdots \tau_{L-1})$ in
the steady state of configuration $\{ \tau_1, \cdots \tau_{L-1} \}$
 (where $\tau_{i}=1$  if site $i$ is occupied and $\tau_i=0$ if it is emtpy);
 we can write the sum over possible particle
numbers  in the steady state in eq.(\ref{relat}) as an explicit sum over
all possible configurations
\begin{equation}
\sum_{m=1}^L \ P_L (m) \ {1 \over q^{m-1}} =
\sum_{\tau_{1}, \cdots\tau_{L-1}}
{ w(\tau_{1},\cdots,\tau_{L-1}) \over  q^{\tau_1+\cdots+\tau_{L-1}}}
\end{equation}
Using the identity $q^{1-\tau_1+...+1-\tau_{L-1}}=
\prod_{j=1}^{L-1}[1+(q-1)(1-\tau_j)]$,
 expanding the product and regrouping
the different terms, we obtain
%\begin{eqnarray}
\begin{equation}
\rho_L (q)=\frac{1}{q^{L-1}} \left[\  1 + \sum_{1\leq i<j\leq L} (q-1)^{j-i}
B_{ij}
\ + \ \sum_{1\leq i<j<k<l \leq L} (q-1)^{j-i +l-k} \ B^{(2)}_{i,j,k,l} + \cdots
\right]\\
\label{sum}
\end{equation}
This gives the $\rho_L(q)$ in terms
of the known $B_{i,j}$  and it can be used in this form.
For example, for $L=4$, it gives $\rho_4(q)= (8 q^3+23 q^2+12q+1)/44q^3$
in agreement with what had been obtained by a direct calculation for
 small sizes \cite{der1}.

An equivalent but more compact and
manageable form  can be  obtained by defining  two antisymmetric matrices
$A_{ij}=-A_{ji}=-(q-1)^{j-i} \mathrm{and}\  C_{ij}=-C_{ji}=B_{i,j}\
\mathrm{for}\
1\leq i<j\leq L$.
Eq.(\ref{sum}) can then be rewritten simply as \cite{no1},
\begin{equation}
\rho_L (q)= \frac{1}{q^{L-1}} \sqrt{\mathrm{det}(1+AC)}
\label{det1}
\end{equation}
This expression (\ref{det1}) is our   main
exact result.

We have only been able to compute exactly the determinant of eq.(\ref{det1})
for q=0 (and 1!)\cite{dhp}. For other values of $q$, we could  only
evaluate its asymptotic behavior. We summarize the main elements
of our estimation which will be detailed elsewhere \cite{dhp}. For large
$L$,
 and $i,j$ far from the boundaries, $B_{i,j}$ becomes a  continuous function
$\beta(x,y)$  of $x=i/L$ and $y=j/L$ which satisfies  the Laplace equation (the
continuous version of eq.(\ref{dif})) with the boundary conditions
$\beta(x,x)=1,\ \beta(0,x)=\beta(x,1)=0$. This  is a slowly
varying function except in the two symmetric arbitrary small corners
$0\leq x < y\ll 1$ and
$1-y < 1-x \ll 1 $ where its
asympotic behavior is
\begin{equation}
\beta(x,y) = \beta(1-y,1-x)=\frac{4}{\pi}  \tan^{-1}(x/y),  \ \ \ \mathrm{for}
\ 0<x<y\ll 1
\label{beta}
\end{equation}
These two small neighbourhoods are responsible for the singular behavior of the
determinant
 for large $L$. In  the region $x<y \ll 1$, $\beta(x,y)$ is
of the form $f(x/y)$ and the singular contribution can be obtained
by computing the traces of the power of a Toeplitz matrix
\cite{kac} in the variables $\log(x),\log(y)$. The same is true in the region
 $1-y < 1-x \ll 1$ where $\beta(x,y) \simeq
f[(1-y)/(1-x)]$ (and with the variables $\log(1-x),\log(1-y)$). In this way
we obtain for $1\leq q <2$,
\begin{equation}
\theta(q) =  -\int_{-\infty}^{\infty} \frac{dk}{4\pi}  \
\log \left( 1 - 2 \frac{q-1}{q^2}
\int_0^1 dx\  (x^{ik} + x^{-ik})\   f' (x)  \right)
\label{expo}
\end{equation}
Using (\ref{beta}), this leads to   (\ref{mform}) for
 $\theta(q)$.
When $q>2$, there is an additional term to (\ref{expo}) coming
from an eigenvalue of the matrix $1+AC$ which becomes very small as $L \to
\infty$.
As a consequence, (\ref{expo}) has to be replaced by its analytic continuation
from
the range $1 < q < 2$ and this leads again to (\ref{mform}).

In the whole range of $q$, the result (\ref{mform}) is in excellent agreement
with previous numerical estimates \cite{DBG,Stauf,der1}. For example, for
$q=5$,
the exact prediction (\ref{mform}) gives $\theta(5)=.6928365..
$ which agrees with the numerical estimate $.6928 \pm .0003$ of \cite{der1}.

Our result (\ref{mform}) shows that  systems as simple as the one dimensional
Ising or Potts model at zero temperature exhibit
 rather complicated power law decays very similar to
 other irreversible processes such as reaction-diffusion
\cite{Racz,reactdif} or sequential-parking \cite{Pome} problems.

Result (\ref{mform}) is also reminiscent of
 the exactly known exponents \cite{denN}  of the equilibrium
 Potts model in dimension 2.  It would be interesting
to see whether the methods originally used   for these
equilibrium problems or the more recent conformal theory techniques
could be extended here to rederive (\ref{mform}).

 We have already noted that expressions (\ref{B2ij}),(\ref{Bnij}),(\ref{sum})
and (\ref{det1})
are Pfaffians \cite{meh} which are always present in free fermion problems.
 In fact, another way of solving the coagulation model (with its boundary)
consists in mapping it onto
a free fermions  problem.
 The details of this solution which requires several successive transformations
of the original problem will be given in \cite{dhp}.
As the  problem of the steady state  of the reaction-diffusion
is completely solved, one could  calculate
steady state properties such as
density profiles, correlation   functions. Also, as the whole problem can be
reduced to a free fermion problem \cite{dhp} (through unfortunately
rather complicated transformations), one should be able to calculate all kinds
of unequal time correlations.

Lastly, it is worth noting that the different
scaling behaviors of $\rho_L (q)$ when $q$ varies,  arise in  (\ref{mform})
from
the  asymptotic form of $P_L (m)$ for $L\gg 1$,
$P_L (m) \sim \exp[\log L\; g(m/\log L)]$
where the function $g$ is a Legendre
transform of $2 \theta(q)$  with respect to $\log(q)$ in a way
quite reminiscent of the usual description of multifractal measures
\cite{Parf}. \\

{\bf Acknowledgments}   We thank Deepak Dhar  for
very stimulating discussions while we were visiting the Isaac Newton Institute
at Cambridge, in particular for suggesting the idea of transforming the  model
into a free fermion problem.

%\newpage
 

\begin{thebibliography}{20}
\bibitem{no1} The reader can convince him/herself of the
algebraic equivalence
of eq.(\ref{sum}) and (\ref{det1}) by an explicit computation of
the determinant for systems of small sizes ($L=3,4,...$). The proof, which is
not difficult, will be given in a longer and more detailed
version of this work\cite{dhp}.
\end{thebibliography}
\end{document}